\RequirePackage{lineno}
\documentclass[aps,showpacs,byrevtex,prd, reprint]{revtex4-1}
\usepackage{graphicx}
\usepackage{dcolumn}
\usepackage{bm}
\usepackage{rotating}
\usepackage{epstopdf}
\usepackage{color}
\usepackage{verbatim} 
\usepackage{multirow}
\usepackage[abs]{overpic}
\usepackage{amsmath}
\usepackage{mathrsfs}
\usepackage{amssymb}
\usepackage{subfigure}
\usepackage{xspace}
\usepackage{float}
\usepackage[colorlinks,
            linkcolor=blue,
            anchorcolor=blue,
            citecolor=blue]{hyperref}

\newcommand{\PreserveBackslash}[1]{\let\temp=\\#1\let\\=\temp}
\newcolumntype{C}[1]{>{\PreserveBackslash\centering}p{#1}}
\newcolumntype{R}[1]{>{\PreserveBackslash\raggedleft}p{#1}}
\newcolumntype{L}[1]{>{\PreserveBackslash\raggedright}p{#1}}
\newcommand{\pp}{\pi^+\pi^-}

\newcommand{\EE}{e^+e^-}

\newcommand{\psip}{\psi(3686)}

\newcommand{\piz}{\pi^{0}}
\newcommand{\etap}{\eta^\prime}

\newcommand{\too}{\rightarrow}

\begin{document}
\graphicspath{{figure/}}
\DeclareGraphicsExtensions{.eps,.png,.ps}
\title{\boldmath Study of the $\chi_{cJ}\too\Lambda\bar{\Lambda}\etap$ decays}

\author{
\begin{small}
\begin{center}
M.~Ablikim$^{1}$, M.~N.~Achasov$^{4,c}$, P.~Adlarson$^{79}$, X.~C.~Ai$^{84}$, R.~Aliberti$^{37}$, A.~Amoroso$^{78A,78C}$, Q.~An$^{75,61,a}$, Y.~Bai$^{60}$, O.~Bakina$^{38}$, Y.~Ban$^{48,h}$
% ...
\\
\vspace{0.2cm}
(BESIII Collaboration)\\
\vspace{0.2cm} {\it
$^{1}$ Institute of High Energy Physics, Beijing 100049, People's Republic of China\\
% ...
$^{4}$ Budker Institute of Nuclear Physics SB RAS (BINP), Novosibirsk 630090, Russia\\
% ...
$^{37}$ Johannes Gutenberg University of Mainz, Johann-Joachim-Becher-Weg 45, D-55099 Mainz, Germany\\
$^{38}$ Joint Institute for Nuclear Research, 141980 Dubna, Moscow region, Russia\\
% ...
$^{48}$ Peking University, Beijing 100871, People's Republic of China\\
% ...
$^{60}$ Southeast University, Nanjing 211100, People's Republic of China\\
$^{61}$ State Key Laboratory of Particle Detection and Electronics, Beijing 100049, Hefei 230026, People's Republic of China\\
% ...
$^{75}$ University of Science and Technology of China, Hefei 230026, People's Republic of China\\
% ...
$^{78}$ University of Turin and INFN, (A)University of Turin, I-10125, Turin, Italy; (B)University of Eastern Piedmont, I-15121, Alessandria, Italy; (C)INFN, I-10125, Turin, Italy\\
$^{79}$ Uppsala University, Box 516, SE-75120 Uppsala, Sweden\\
% ...
$^{84}$ Zhengzhou University, Zhengzhou 450001, People's Republic of China\\
\vspace{0.2cm}
$^{a}$ Deceased\\
% ...
$^{c}$ Also at the Novosibirsk State University, Novosibirsk, 630090, Russia\\
% ...
$^{h}$ Also at State Key Laboratory of Nuclear Physics and Technology, Peking University, Beijing 100871, People's Republic of China\\
% ...
}
\end{center}
\vspace{0.4cm}                                                                         
\end{small}
}


\begin{abstract}
Using a data sample of $(2.712\pm0.014)\times10^{9}$ $\psi(3686)$ events collected with the BESIII detector at the BEPCII collider, we investigate the decays $\chi_{cJ} \rightarrow \Lambda \bar{\Lambda} \etap$ for $J=0,~1,~2$ via the radiative transition $\psi(3686) \rightarrow \gamma \chi_{cJ}$.
The decays $\chi_{c0,2}\too\Lambda\bar{\Lambda}\etap$ are observed for the first time, with statistical significances of 6.7$\,\sigma$ and 6.4$\,\sigma$, respectively.
Evidence for the decay $\chi_{c1}\too\Lambda\bar{\Lambda}\etap$ is found with a statistical significance of 3.3$\,\sigma$.
The corresponding branching fractions are measured to be $\mathscr{B}(\chi_{c0}\too\Lambda\bar{\Lambda}\etap)=(7.56\pm1.42\pm0.90)\times10^{-5}$, $\mathscr{B}(\chi_{c1}\too\Lambda\bar{\Lambda}\etap)=(1.54\pm0.51\pm0.16)\times10^{-5}$, and $\mathscr{B}(\chi_{c2}\too\Lambda\bar{\Lambda}\etap)=(3.03\pm0.61\pm0.29)\times10^{-5}$, where the first uncertainties are statistical and the second systematic. No significant excited $\Lambda$ baryon states or $\Lambda\bar{\Lambda}$ near-threshold enhancements are observed.
\end{abstract}


\maketitle

\section{INTRODUCTION}

The $\chi_{cJ}~(J=0,~1,~2)$ mesons are the $P$-wave spin-triplet charmonium states in the quark model.  
Compared with the extensively studied decays of the $S$-wave charmonia $J/\psi$ and $\psi'$, the decays of $\chi_{cJ}$ remain less thoroughly investigated both experimentally and theoretically.  
Their direct production in $e^{+}e^{-}$ annihilation is strongly suppressed since they involve processes with two virtual photons or neutral currents; to date, only the $\chi_{c1}$ state has been observed at BESIII with a significance of $5.1\sigma$~\cite{ee_chic1} in these processes.  
An efficient production mechanism for the $\chi_{cJ}$ states is provided by the radiative transitions $\psi'\rightarrow\gamma\chi_{cJ}$, whose branching fractions are approximately $10\%$ for each $\chi_{cJ}$ state.  
This enables precision studies of $\chi_{cJ}$ decays with a large $\psi'$ data sample.

Decays of $\chi_{cJ}$ mesons into baryon–antibaryon–meson ($B\bar{B}M$) final states offer a unique opportunity to investigate $B\bar{B}$ threshold behaviour and excited baryon states in the $BM$ system, where $B$ and $\bar{B}$ denote a baryon and its antibaryon, respectively, and $M$ represents a meson.  
Recently, evidence for a resonance $\Lambda^{*}$ was reported in the invariant-mass spectrum of $\Lambda(\bar{\Lambda})$ and $\omega$ in the decay $\psi(3686)\rightarrow\Lambda\bar{\Lambda}\omega$~\cite{lam2omega}.  
Although $\Lambda\bar{\Lambda}$ mass-threshold enhancements have been observed in $e^{+}e^{-}\rightarrow\Lambda\bar{\Lambda}\phi$~\cite{lam2phi} and $e^{+}e^{-}\rightarrow\Lambda\bar{\Lambda}\eta$~\cite{lam2eta}, significant discrepancies in the magnitude of the enhancements are evident between the two channels.  
The $\chi_{cJ}\rightarrow B\bar{B}M$ decay channels ($\chi_{cJ}\rightarrow\Lambda\bar{\Lambda}\eta$~\cite{chicj1}, $\Lambda\bar{\Lambda}\omega$~\cite{chicj2}, $\Lambda\bar{\Lambda}\phi$~\cite{chicj3}, and $\Sigma^{+}\bar{\Sigma}^{-}\eta$~\cite{chicj4}) have been exploited to search for such enhancements, but no conclusive evidence has yet been obtained.

In this paper we investigate the decays $\chi_{cJ}\rightarrow\Lambda\bar{\Lambda}\eta^{\prime}$ using a data sample of $(2.712\pm0.014)\times10^{9}$ $\psi(3686)$ events~\cite{psipdata} collected with the BESIII detector.  
The $\Lambda$ and $\bar{\Lambda}$ baryons are reconstructed via $\Lambda\rightarrow p\pi^{-}$ and $\bar{\Lambda}\rightarrow\bar{p}\pi^{+}$, respectively.  
The $\eta^{\prime}$ meson is reconstructed through two decay modes: $\eta^{\prime}\rightarrow\gamma\pi^{+}\pi^{-}$ (Mode I) and $\eta^{\prime}\rightarrow\eta\pi^{+}\pi^{-}$ with $\eta\rightarrow\gamma\gamma$ (Mode II).  
The branching fractions of $\chi_{cJ}\rightarrow\Lambda\bar{\Lambda}\eta^{\prime}$ are measured for the first time.  
In addition, possible $\Lambda^{*}$ excited states and near-threshold enhancement in the $\Lambda\bar{\Lambda}$ system are investigated.

\section{BESIII DETECTOR AND MONTE CARLO SIMULATION}

The BESIII detector~\cite{Ablikim:2009aa} records symmetric $e^+e^-$ collisions 
provided by the BEPCII storage ring~\cite{Yu:IPAC2016-TUYA01}
in the center-of-mass energy range from 1.84 to 4.95~GeV,
with a peak luminosity of $1.1 \times 10^{33}\;\text{cm}^{-2}\text{s}^{-1}$ 
achieved at $\sqrt{s} = 3.773\;\text{GeV}$. 
BESIII has collected large data samples in this energy region~\cite{Ablikim:2019hff, EcmsMea, EventFilter}. The cylindrical core of the BESIII detector covers 93\% of the full solid angle and consists of a helium-based
 multilayer drift chamber~(MDC), a time-of-flight
system~(TOF), and a CsI(Tl) electromagnetic calorimeter~(EMC),
which are all enclosed in a superconducting solenoidal magnet
providing a 1.0~T magnetic field.
The solenoid is supported by an
octagonal flux-return yoke with resistive plate counter muon
identification modules interleaved with steel. 
The charged-particle momentum resolution at $1~{\rm GeV}/c$ is
$0.5\%$, and the 
${\rm d}E/{\rm d}x$
resolution is $6\%$ for electrons
from Bhabha scattering. The EMC measures photon energies with a
resolution of $2.5\%$ ($5\%$) at $1$~GeV in the barrel (end cap)
region. The time resolution in the plastic scintillator TOF barrel region is 68~ps, while
that in the end cap region was 110~ps. The end cap TOF
system was upgraded in 2015 using multigap resistive plate chamber
technology, providing a time resolution of
60~ps, which benefits 83\% of the data used in this analysis~\cite{etof}.

Monte Carlo (MC) simulated data samples produced with a {\sc geant4}-based~\cite{geant4} software package, which includes the geometric description of the BESIII detector and the detector response, are used to determine detection efficiencies and to estimate backgrounds. The simulation models the beam energy spread and initial state radiation (ISR) in the $e^+e^-$
annihilations with the generator {\sc kkmc}~\cite{KKMC}. The inclusive MC sample includes the production of the $\psi(3686)$ resonance, the ISR production of the $J/\psi$, and the continuum processes incorporated in {\sc kkmc}~\cite{KKMC}. All particle decays are modeled with {\sc
evtgen}~\cite{ref:evtgen} using branching fractions either taken from the Particle Data Group (PDG)~\cite{pdg}, when available, or otherwise estimated with {\sc lundcharm}~\cite{ref:lundcharm}.
Final state radiation from charged final state particles is incorporated using the {\sc photos} package~\cite{photos}.

The signal MC events of the $\psi(3686)\too\gamma \chi_{cJ}$ decays follow the angular distribution of $1+\lambda \cos^{2}{\theta_{\gamma}}$, where $\theta_{\gamma}$ denotes the polar angle of the radiative photon, and $\lambda$ takes the values 1, $-1/3$, and 1/13 for $J=0$, 1, and 2, respectively~\cite{gammaangle}. For subsequent decays, the processes $\chi_{cJ}\too\Lambda\bar{\Lambda}\etap$ are modeled using a phase-space (PHSP) model. The decay $\eta^\prime\too\gamma\pi^{+}\pi^{-}$ is implemented via the DIY model~\cite{etapgam2pi} incorporating the $\rho-\omega$ interference effects and the box anomaly mechanism. The decay $\eta^\prime\too\eta\pi^{+}\pi^{-}$ is generated according to the Dalitz distribution of the Ref.~\cite{etapeta2pi} with the subsequent decay of $\eta\too\gamma\gamma$ following a PHSP model. 
The processes of $\Lambda\too p\pi^{-}$ and $\bar{\Lambda}\too\bar{p}\pi^{+}$ are generated by the HypWK model~\cite{ref:evtgen}, which is constructed for spin-$\frac{1}{2}$ hyperon and its anti-hyperon non-leptonic decays including
parity violation effects. 

\section{EVENT SELECTION}

Candidate events are required to have the final states $p\bar{p}\pi^{+}\pi^{-}\pi^{+}\pi^{-}2\gamma$ (Mode I) and $p\bar{p}\pi^{+}\pi^{-}\pi^{+}\pi^{-}3\gamma$ (Mode II).  Charged tracks detected in the MDC are required to be within a polar angle range of $(|\!\cos\theta|<0.93)$, where $\theta$ is defined with respect to the $z$-axis, which is the symmetry axis of the MDC. At least three positive and three negative tracks are required for each mode.  
$\Lambda$ and $\bar{\Lambda}$ candidates are reconstructed by pairing oppositely charged tracks with proton and pion mass hypotheses, respectively. Each pair is subjected to a primary vertex fit followed by a secondary vertex fit; the latter is applied to reconstruct long-lived resonances~\cite{secondfit}.
The optimal combination is selected by minimizing the mass deviation, defined as $\Delta=\sqrt{(M(p\pi^{-})-m_{\Lambda})^{2}+(M(\bar{p}\pi^{+})-m_{\bar{\Lambda}})^{2}}$, where $M(p\pi^{-})$ and $M(\bar{p}\pi^{+})$ are the invariant masses of $p\pi^{-}$ and $\bar{p}\pi^{+}$, respectively, and $m_{\Lambda(\bar{\Lambda})}$ denotes the $\Lambda~(\bar{\Lambda})$ nominal mass~\cite{pdg}. Tracks not associated with $\Lambda$ or $\bar{\Lambda}$ are required to originate from the interaction point (IP) within $10~\text{cm}$ along the $z$ axis and $1~\text{cm}$ in the transverse plane.  
Exactly two such tracks with zero net charge are retained and assigned the pion mass hypothesis; they are subjected to a vertex fit to confirm their common origin.

Photon candidates are identified from showers in the EMC. The deposited energy of each shower must be more than 25 MeV in the barrel region ($|\!\cos\theta|<0.80$) and more than 50 MeV in the end cap region ($0.86<|\!\cos\theta|<0.92$). To exclude showers that originate from charged tracks, the angle subtended by the EMC shower and the position of the closest charged track at the EMC must be greater than 10 degrees as measured from the IP. To suppress electronic noise and showers unrelated to the event, the difference between the EMC time and the event start time must be within [0, 700] ns. At least two (three) photons are required for Mode I (Mode II).

\begin{figure}[htbp]
\begin{center}
\begin{overpic}[width=0.40\textwidth, trim=35 0 20 0]{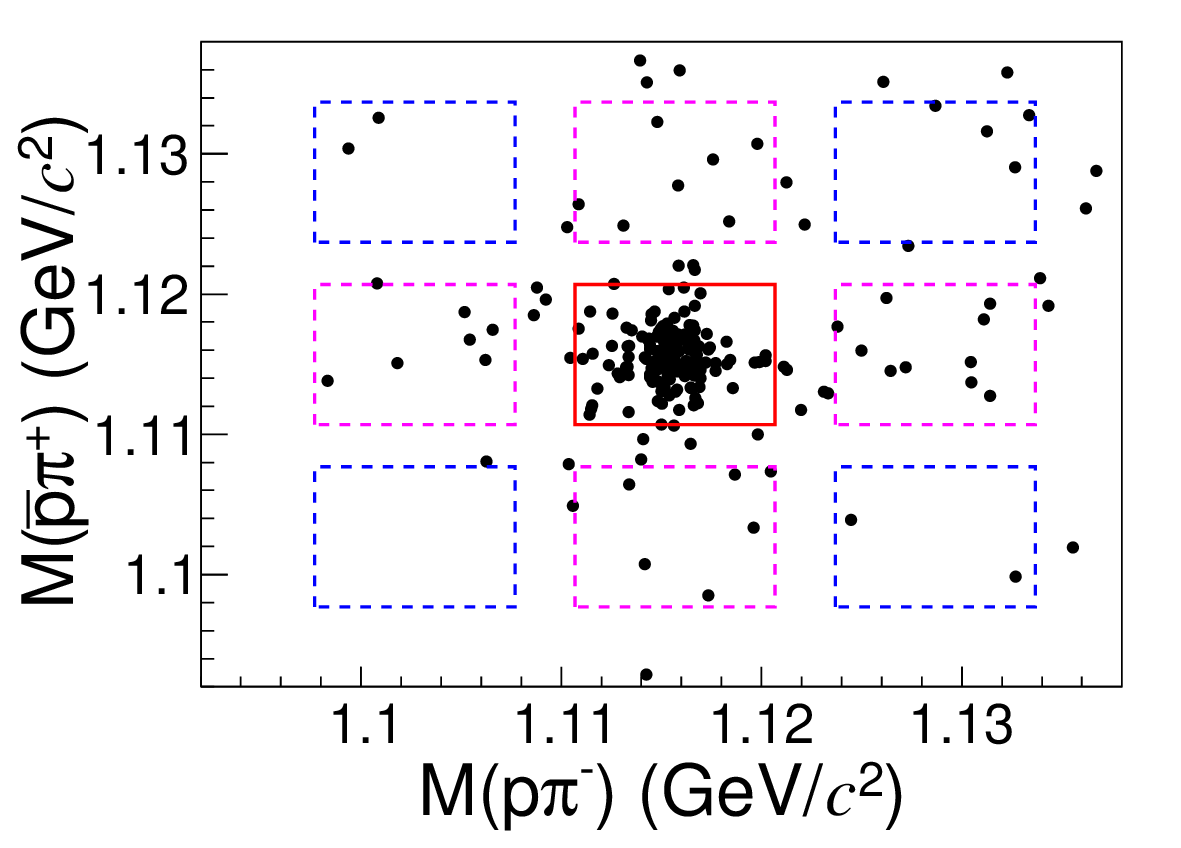}
\put(27,141){\large{(I)}}
\end{overpic}
\begin{overpic}[width=0.40\textwidth, trim=35 20 20 0]{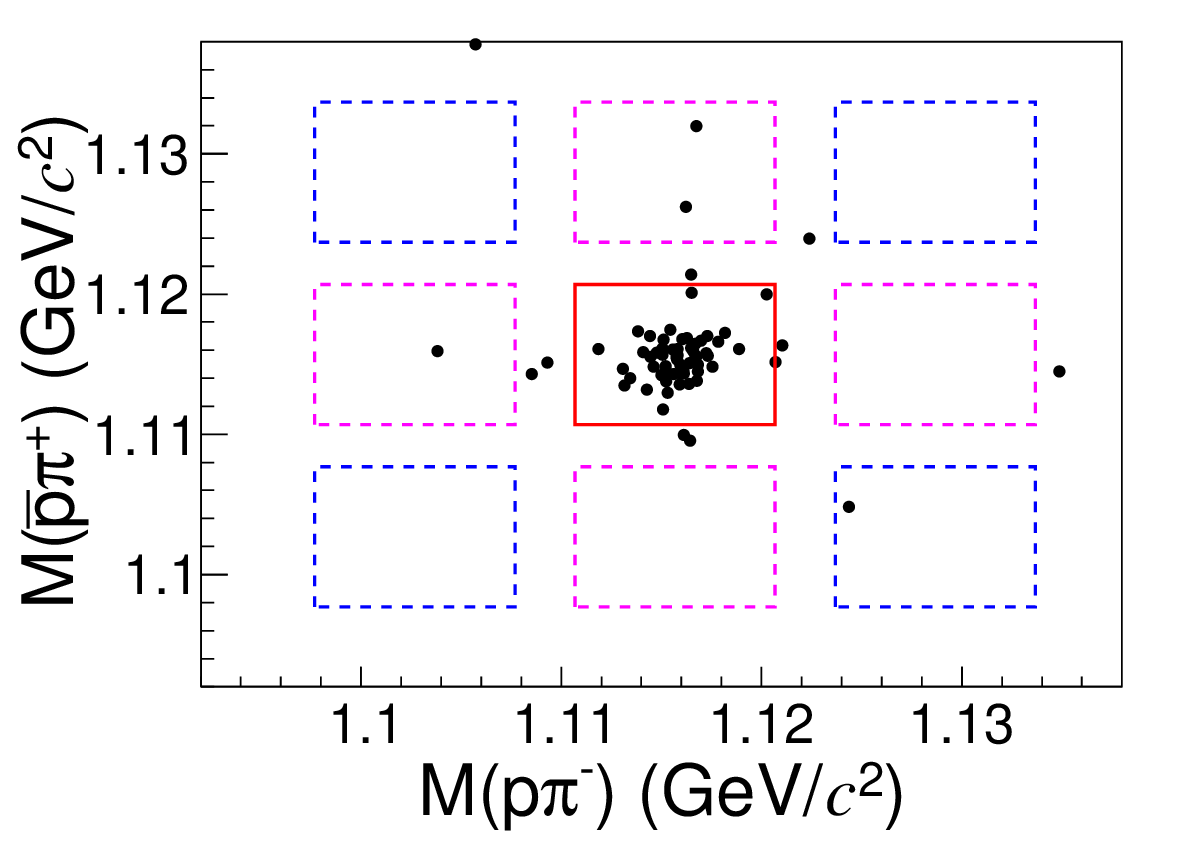}
\put(27,133){\large{(II)}}
\end{overpic}
\caption{The scatter plots of $M(\bar{p}\pi^{+})$ versus $M(p\pi^{-})$ for Mode I and Mode II from the survived candidates in data, where the red solid box is the 2-D $\Lambda\bar{\Lambda}$ signal region, and the pink and blue dashed boxes are the 2-D $\Lambda\bar{\Lambda}$ sideband regions.}
\label{fig:lambda_2D}
\end{center}
\end{figure}

To improve mass resolution and suppress background, a four-constraint (4C) kinematic fit is applied to Mode I, constraining the total four-momentum of the final-state particles to that of the colliding beams. When more than two photon candidates are present, the combination with the least $\chi^{2}_{4\rm C}$ is selected, satisfying $\chi^{2}_{\rm 4C}<18$. The photon originating from the $\etap$ decay is identified as the one that minimises $|M(\gamma\pi^{+}\pi^{-})-m_{\etap}|$, where $M(\gamma\pi^{+}\pi^{-})$ is the reconstructed invariant mass of $\gamma\pi^{+}\pi^{-}$, and  $m_{\etap}$ is the $\etap$ nomial mass~\cite{pdg}. For Mode II, a five-constraint (5C) kinematic fit is implemented: a 4C kinematic fit together with an additional constraint on the two photons from the $\eta$ candidate to the $\eta$ nominal mass~\cite{pdg}.  When more than three photon candidates are present, the combination with the least $\chi^{2}_{5\rm C}$ is chosen, requiring $\chi^{2}_{5\rm C}<53$. The $\chi^{2}$ requirements for both modes are optimized by maximizing the figure of merit $ S/\sqrt{S+B}$, where $S(B)$ is the number of signal (background) events from the normalized signal (inclusive) MC sample. In addition, the criterion $\chi^{2}_{4{\rm C},n\gamma}<\chi^{2}_{4{\rm C},(n+1)\gamma}$ is applied for both modes, where $n$ is the expected number of photons in the signal process.
The $\Lambda~(\bar{\Lambda})$ signal region is defined as $|M(p\pi^{-})-m_{\Lambda}|<5 ~{\rm MeV}/c^{2}$ ($|M(\bar{p}\pi^{+})-m_{\bar{\Lambda}}|<5~{\rm MeV}/c^{2}$), with the additional requirement that the decay length of the $\Lambda~(\bar{\Lambda})$ candidate must be larger than zero. Figure~\ref{fig:lambda_2D} shows the distributions of $M(\bar{p}\pi^{+})$ versus $M(p\pi^{-})$ for Mode I and Mode II for the surviving candidates in the data. The box outlined by red solid lines is the two-dimensional (2-D) $\Lambda\bar{\Lambda}$ signal region, and the boxes outlined by pink and blue dashed lines are 2-D $\Lambda\bar{\Lambda}$ sideband regions. A clear $\Lambda\bar{\Lambda}$ pair signal is found in the distribution. The events in the 2-D $\Lambda\bar{\Lambda}$ sideband regions are used to estimate the non-$\Lambda\bar{\Lambda}$ background in the 2-D $\Lambda\bar{\Lambda}$ signal region. 

\begin{figure}[htbp]
\begin{center}
\begin{overpic}[width=0.40\textwidth, trim=35 0 20 0]{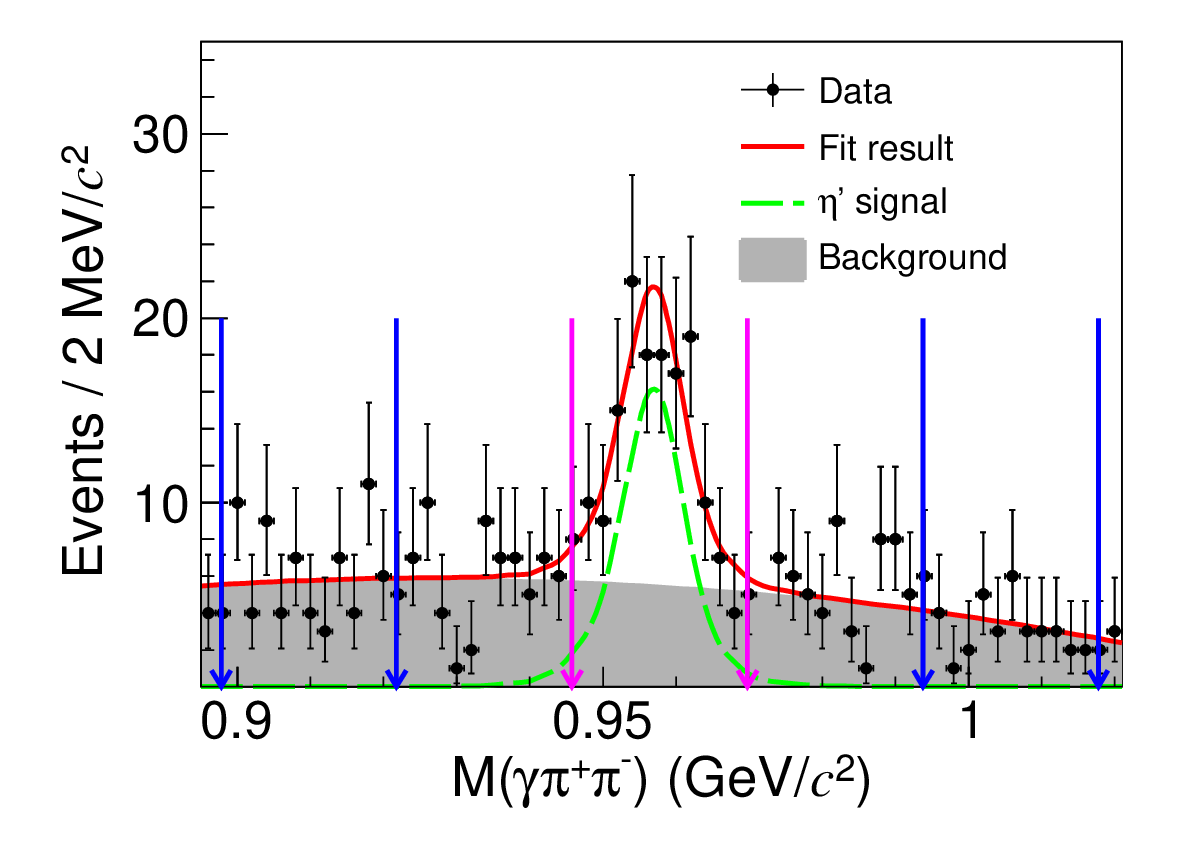}
\put(45,130){\large{(I)}}
\end{overpic}
\begin{overpic}[width=0.40\textwidth, trim=35 20 20 0]{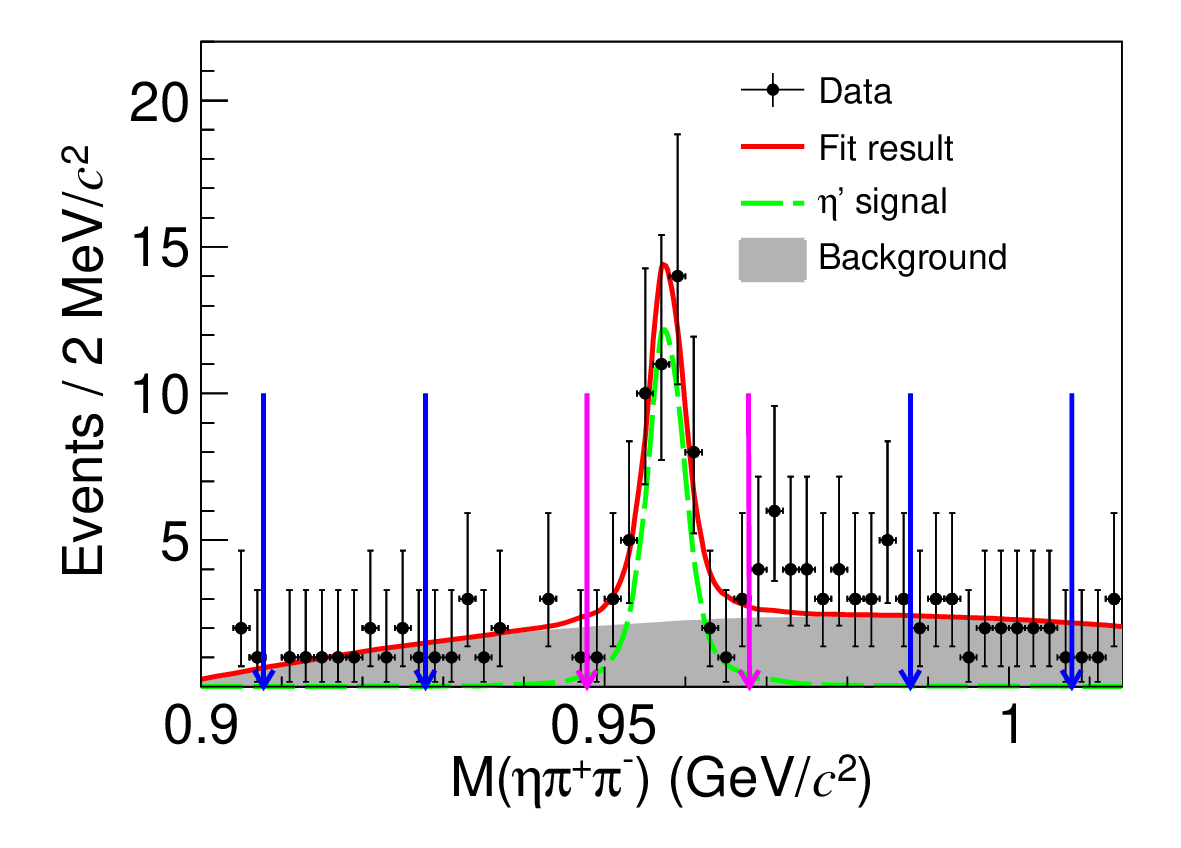}
\put(45,125){\large{(II)}}
\end{overpic}
\caption{The distributions of $M(\gamma\pp)$ for Mode I and $M(\eta\pp)$ for Mode II. The pink arrows mark the $\etap$ signal region, and the two pairs of blue arrows mark the $\etap$ sideband regions. The dots with error bars are data, the red solid lines are the fit results, the green dashed lines represent the signal shapes, and the grey shaded histograms constitute the smooth backgrounds.}
\label{fig:fit_metap}
\end{center}
\end{figure}

For Mode I, the directly emitted photon is labelled $\gamma_{1}$ and that from the $\eta^{\prime}$ decay $\gamma_{2}$. After applying the aforementioned selection criteria, three types of background processes remain, the $\Sigma^{0}(\bar{\Sigma}^{0})$-related backgrounds, the $J/\psi$-related backgrounds, and the $\piz$-related backgrounds. To suppress the $\Sigma^{0}(\bar{\Sigma}^{0})$-related backgrounds, we require the invariant mass of the $\Lambda/\bar{\Lambda}$ combined with either photon candidate to satisfy $M(\Lambda(\bar{\Lambda})\gamma_{1, 2})>1.2~{\rm GeV}/c^{2}$. The $J/\psi$-related backgrounds are mitigated by excluding events where the recoil mass of the $\pp$ system and $\gamma_{1}\gamma_{2}$ system fall within the $J/\psi$ mass window, $|RM(\pi^{+}\pi^{-})-m_{J/\psi}|>10~{\rm MeV}/c^{2}$ and $|RM(\gamma_{1}\gamma_{2})-m_{J/\psi}|>26~{\rm MeV}/c^{2}$. The $\piz$-related backgrounds are suppressed by vetoing the $\piz$ mass region in the $\gamma_{1}\gamma_{2}$ invariant mass distribution, $|M(\gamma_{1}\gamma_{2})-m_{\piz}|>14~{\rm MeV}/c^{2}$. The $\etap$ signal region is set to be $[m_{\etap}-12,~m_{\etap}+12]~{\rm MeV}/c^{2}$, with $\etap$ sideband regions $[m_{\etap}-60,~m_{\etap}-36]\cup[m_{\etap}+36,~m_{\etap}+60]~{\rm MeV}/c^{2}$ used to estimate the non-$\etap$ background in the $\etap$ signal region. For Mode II, the requirement of $M(\Lambda(\bar{\Lambda})\gamma)>1.2~{\rm GeV}/c^{2}$ is used to reject the $\Sigma^{0}(\bar{\Sigma}^{0})$-related backgrounds. The $\etap$ signal region is tightened to $[m_{\etap}-10,~m_{\etap}+10]$ MeV/$c^{2}$, with the $\etap$ sideband regions set to be $[m_{\etap}-50,~m_{\etap}-30]\cup[m_{\etap}+30,~m_{\etap}+50]~{\rm MeV}/c^{2}$. Figure~\ref{fig:fit_metap} shows the distributions of $M(\gamma\pp)$ for Mode I and $M(\eta\pp)$ for Mode II in data, with the $\etap$ signal and sideband regions indicated. 

\section{BRANCHING FRACTIONS}

\begin{figure*}[htbp]
\begin{center}
\begin{overpic}[width=0.80\textwidth, trim=35 20 20 20]{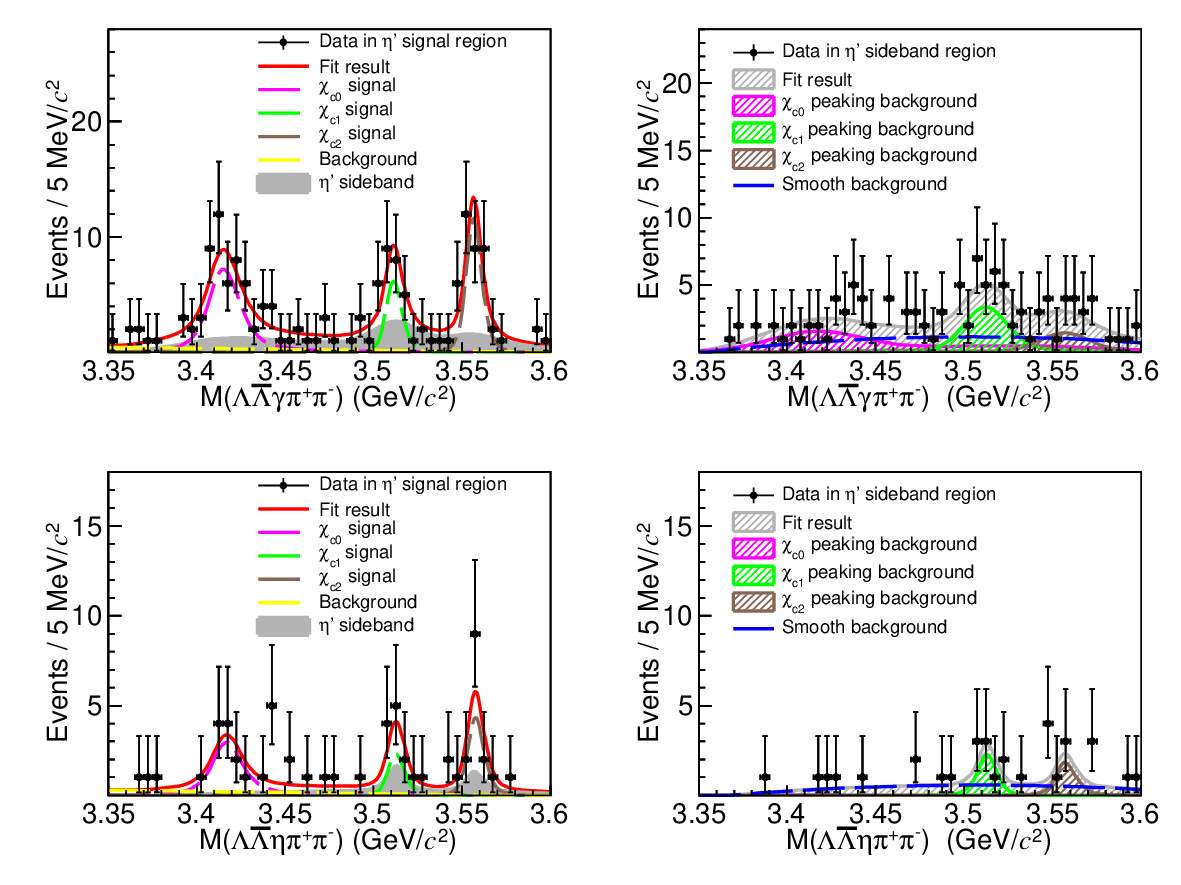}
\put(27,288){\large{(A)}}
\put(370,288){\large{(B)}}
\put(27,118){\large{(C)}}
\put(370,118){\large{(D)}}
\end{overpic}
\caption{Simultaneous fit to the $M(\Lambda\bar{\Lambda}\gamma\pp)$ (up) and $M(\Lambda\bar{\Lambda}\eta\pp)$ (bottom) distributions of the accepted candidate events in the $\etap$ (left) signal and (right) sideband regions. The dots with error bars are the data, the grey shaded histograms constitute the total contribution from the $\etap$ sideband, and the pink, green, and brown shaded histograms are the peaking background from $\chi_{c0}$, $\chi_{c1}$, and $\chi_{c2}$, respectively. The solid red lines are the fit results, and the pink, green, and brown dashed lines represent the $\chi_{c0}$, $\chi_{c1}$, and $\chi_{c2}$ signal shapes, respectively.}
\label{fig:fit_mchicj}
\end{center}
\end{figure*}

The branching fractions of $\chi_{cJ}\too\Lambda\bar{\Lambda}\etap$ are determined using a simultaneous unbinned maximum-likelihood fit to the distributions of the invariant masses of $M(\Lambda\bar{\Lambda}\gamma\pp)$ for Mode I and $M(\Lambda\bar{\Lambda}\eta\pp)$ for Mode II.  
The fit is performed in both the $\etap$ signal and sideband regions. In the simultaneous fit, a common branching fraction $\mathscr{B}(\chi_{cJ}\too\Lambda\bar{\Lambda}\etap)$ is shared between the two modes. Furthermore, the parameters describing the shape of the non-$\etap$ background in the $\etap$ signal region are determined from and shared with the shape of the events in the $\etap$ sideband regions.
For Mode I, the events in the $\etap$ sideband regions are used to estimate the non-$\etap$ background in the $\etap$ signal region. 
The non-$\Lambda\bar{\Lambda}$ background is found to be negligible using the events in the 2-D $\Lambda\bar{\Lambda}$ sideband regions in Fig.\ref{fig:lambda_2D}.
Figures~\ref{fig:fit_mchicj} (A) and \ref{fig:fit_mchicj} (B) show the distributions of $M(\Lambda\bar{\Lambda}\gamma\pp)$ in the $\etap$ signal region and sideband regions, respectively. Peaking backgrounds in the $\etap$ sideband regions are modeled by three $\chi_{cJ}\too\Lambda\bar{\Lambda}\gamma\pp$ PHSP MC shapes plus a second-order Chebyshev polynomial, as shown in Fig.~\ref{fig:fit_mchicj} (B). In the $\etap$ signal region, the $\chi_{cJ}$ signal shapes are described by the corresponding MC shapes, and the smooth background is modeled by a first-order Chebyshev polynomial. 
The non-$\etap$ background is modeled with the same parameterized shapes in the $\etap$ sidebands, as presented in the grey shaded histograms in Figs.~\ref{fig:fit_mchicj} (A) and \ref{fig:fit_mchicj} (B). The normalization factor between the $\etap$ signal and sideband regions is determined from a fit to the $M(\gamma\pp)$ distribution in Fig.~\ref{fig:fit_metap} (I), where the MC shape is used to describe the signal shape, and a second-order Chebyshev polynomial function is used to model the background shape. The normalization factor is the ratio of background yields between the $\etap$ signal and sideband regions. An identical procedure is applied to Mode II, as shown in Figs.~\ref{fig:fit_mchicj} (C) and \ref{fig:fit_mchicj} (D), using $M(\Lambda\bar{\Lambda}\eta\pp)$ as the observable and the sideband normalization determined from the $M(\eta\pp)$ distribution in Fig.~\ref{fig:fit_metap} (II).

The number of signal events for each decay mode is expressed by
\begin{equation}
\label{eq:1}
\begin{aligned}
&N^{\rm sig}_{X}=N^{\rm{tot}}\cdot{\mathscr{B}(\psi(3686)\too \gamma\chi_{cJ}})\cdot\mathscr{B}(\chi_{cJ}\too\Lambda\bar{\Lambda}\etap)  \\
& \cdot{\mathscr{B}(\Lambda\too p\pi^{-})}
\cdot{\mathscr{B}(\bar{\Lambda}\too\bar{p}\pi^{+})}\cdot{\mathscr{B}(\etap\too X)}\cdot\epsilon_{X},
\end{aligned}
\end{equation}
where $X$ denotes the $\etap$ decay mode($\gamma\pi^{+}\pi^{-}$ or $\eta\pi^{+}\pi^{-}$ with $\eta\rightarrow\gamma\gamma$), and $\epsilon_{X}$ is the corresponding detection efficiency from a mixed MC sample described below.  The branching fractions of the intermediate decays are taken from the PDG~\cite{pdg}. The fit projections for both decay modes are shown in Fig.~\ref{fig:fit_mchicj}. The statistical significances of the $\chi_{cJ}\rightarrow\Lambda\bar{\Lambda}\eta^{\prime}$ signals are $7.8\sigma$, $3.5\sigma$, and $7.6\sigma$ for $J=0$, $1$, and $2$, respectively, determined from the change in log-likelihood and the number of degrees of freedom with and without the signal component.  
The numerical fit results are listed in Table~\ref{tab:result}.

\begin{table*}[htbp]
\centering
\caption{Measured branching fractions for $\chi_{cJ}\too\Lambda\bar{\Lambda}\etap$, where the first uncertainties are statistical, the second systematic, S is the significance, and $R_{X(2264)}$ denotes the ratio of the $X(2264)$-intermediate to non-resonant contribution.  }
\begin{tabular}{cccccccccc}
\hline
\hline
    Decay mode    \ \  & \ \  \ \ $\mathcal{B}(\times10^{-5})$ \  \  \  \  \  & $R_{X(2264)}$ (\%) &  \  \  \ Efficiency (I) (\%)    \  \  \  &    \  \   Efficiency (II) (\%) & \  \  \  \ S($\sigma$) \  \  \  \  &  \  \  \  \  \ S include syst. ($\sigma$)  \\
\hline
   $\chi_{c0}\too\Lambda\bar{\Lambda}\etap$  \  \  &  $7.56\pm1.42\pm0.90$ & 41.3 & 1.68 & 1.16 & 7.8 & 6.7  \\
    $\chi_{c1}\too\Lambda\bar{\Lambda}\etap$  \  \  &  $1.54\pm0.51\pm0.16$  & 9.2 & 3.14 & 2.13 & 3.5 & 3.3 \\
   $\chi_{c2}\too\Lambda\bar{\Lambda}\etap$  \  \   &  $3.03\pm0.61\pm0.29$ & 4.8 & 3.04 &    2.12 & 7.6 & 6.4\\

\hline
\hline
\end{tabular}
\label{tab:result}
\end{table*}

\begin{figure*}[htbp]
\begin{center}
\begin{overpic}[width=0.72\textwidth, trim=50 0 50 0]{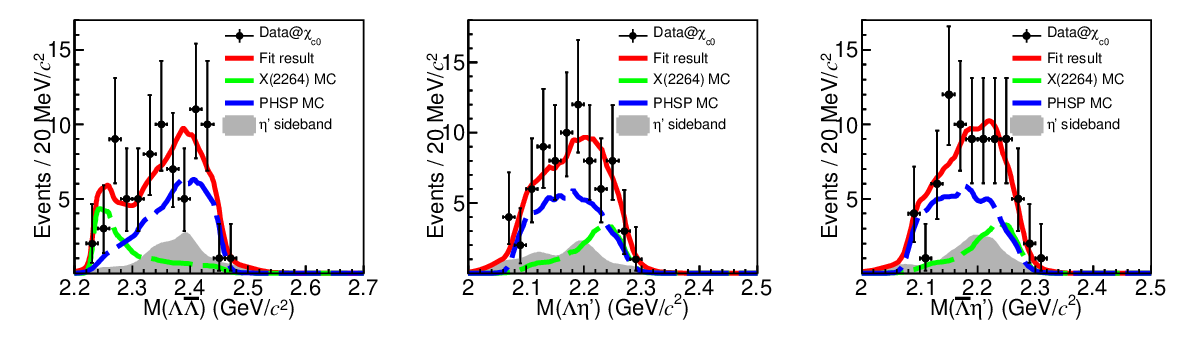}
\end{overpic}
\begin{overpic}[width=0.72\textwidth, trim=50 0 50 0]{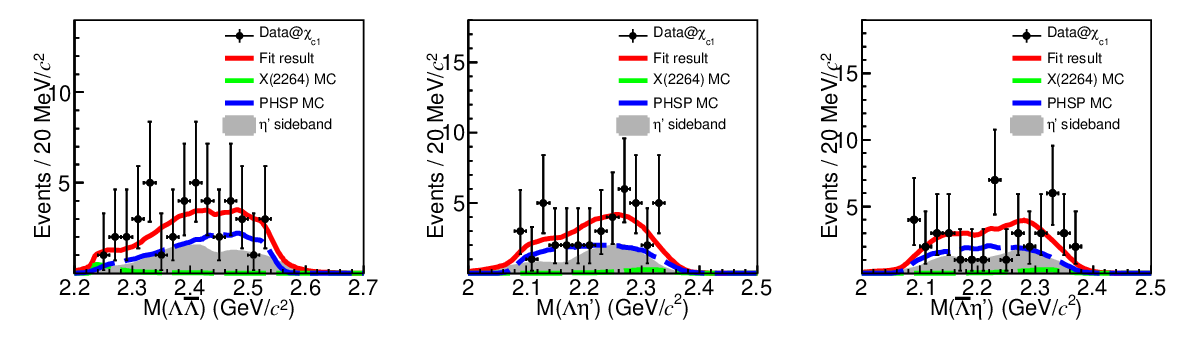}
\end{overpic}
\begin{overpic}[width=0.72\textwidth, trim=50 20 50 0]{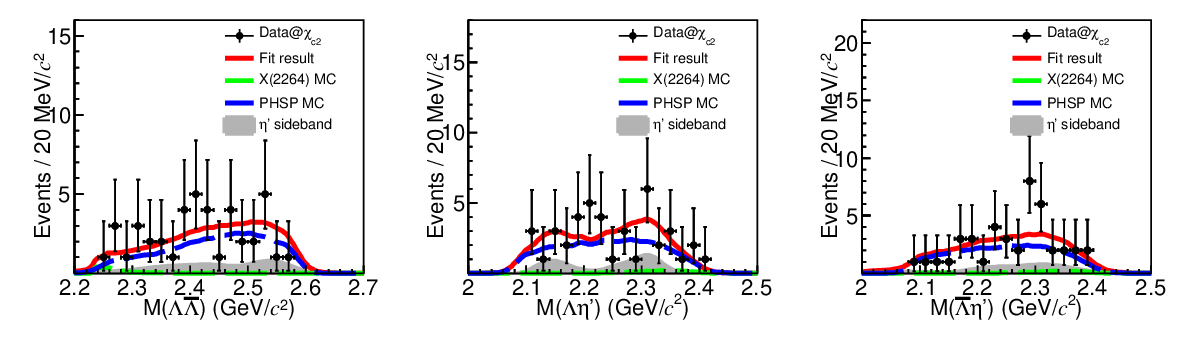}
\end{overpic}
\caption{The projections on two-body invariant mass of the 3-D fit for each $\chi_{cJ}$ decay. The dots with error bars are the data by combining two $\etap$ decay modes, the red solid lines are the fit results, the grey shaded histograms denote the contributions from the $\etap$ sideband, and the green and blue dashed lines represent the $X(2264)$-intermediate state process and non-resonant process, respectively. Events have been required to be within individual $\chi_{cJ}$ signal regions. }
\label{fig:chicj_3D_fit}
\end{center}
\end{figure*}

To evaluate the signal efficiency, a mixed MC sample is generated accounting for possible intermediate states. 
In particular, a near-threshold enhancement in the $\Lambda\bar{\Lambda}$ invariant-mass spectrum observed in $e^{+}e^{-}\rightarrow\phi\Lambda\bar{\Lambda}$~\cite{lam2phi} is considered. 
To evaluate its contribution, we perform a three-dimensional (3-D) unbinned maximum-likelihood fit to the distributions of $M(\Lambda\bar{\Lambda})$, $M(\Lambda\etap)$, and $M(\bar{\Lambda}\etap)$. The mass and width of the $\Lambda\bar{\Lambda}$ enhancement are fixed at 2264 MeV/$c^{2}$ and 73 MeV, respectively~\cite{lam2phi}. We denote this state as $X(2264)$ in the following study. The background is modeled using events from the $\etap$ sidebands with fixed shape and normalisation.  The signal is described as the sum of the $X(2264)$-intermediate state and the non-resonant process, each represented by its respective PHSP MC shapes. Because of limited statistics in each $\eta^{\prime}$ decay mode, the data from both modes are combined. The region of [3.390, 3.445] GeV/$c^{2}$ is used for $\chi_{c0}$, [3.495, 3.525] GeV/$c^{2}$ for $\chi_{c1}$, and [3.540, 3.570] GeV/$c^{2}$ for $\chi_{c2}$. Figure~\ref{fig:chicj_3D_fit} shows the results from the 3-D fit. From the best-fit parameters, the ratio $R_{X(2264)}$ of the $X(2264)$-intermediate to non-resonant contribution is extracted and used to mix the corresponding MC samples.  
The efficiencies obtained from the mixed samples and the $X(2264)$ fractions are listed in Table~\ref{tab:result}.

\section{SYSTEMATIC UNCERTAINTY}

The systematic uncertainties in the measurements of branching fractions are listed in Table~\ref{tab:summererror}. These uncertainties are categorized into correlated and uncorrelated sources between the two $\etap$ decay modes. The systematic uncertainties from the tracking for $\pi^\pm$, the photon detection, the $\Lambda$ reconstruction, the $\bar{\Lambda}$ reconstruction, the kinematic fit, and the signal MC model are the correlated items in each $\etap$ decay mode. The combined correlated uncertainty for the two decay channels is calculated by 
\begin{equation}
\label{eq:2}
\delta_{\rm corr} = \frac{\Sigma_{i=1}^{2}(\delta_{i}\times\mathscr{B}_{i}\times\epsilon_{i})}{\Sigma_{i=1}^{2}(\mathscr{B}_{i}\times\epsilon_{i})}, 
\end{equation}
where $\delta_{i}$, $\mathscr{B}_{i}$, and $\epsilon_{i}$ represent the systematic uncertainty, branching fraction, and detection efficiency for the channel $i$, respectively. The systematic uncertainties from the $\etap$ mass window, the wrong photon combination, the background veto, and the branching fraction of $\etap\too X$ are uncorrelated items. The average of the two decay channels is calculated as
\begin{equation}
\label{eq:3}
\delta_{\rm uncorr} = \frac{\sqrt{\Sigma_{i=1}^{2}(\delta_{i}\times\mathscr{B}_{i}\times\epsilon_{i})^{2}}}{\Sigma_{i=1}^{2}(\mathscr{B}_{i}\times\epsilon_{i})}.
\end{equation}

\begin{table*}[htbp]
  \centering
  \caption{Relative systematic uncertainties (\%) on the branching-fraction measurements. The dash (-) indicates that the source is not relevant for that decay. }
  \label{tab:summererror}
  \begin{minipage}[t]{1\textwidth}
  \centering
  \begin{tabular}{cccccccc}
  \hline
  \hline
    &    \multicolumn{2}{c}{$\chi_{c0}$} \ \  &    \multicolumn{2}{c}{$\chi_{c1}$} \ \  &       \multicolumn{2}{c}{$\chi_{c2}$}  \ \ \    \\
  \hline
 Source  \ \  &   Mode I &  \ \    Mode II\ \  \  \ \  &    Mode I &  \ \    Mode II\ \   &    Mode I &  \ \    Mode II\   \\
  \hline
  Tracking for $\pi^\pm$             &  2.0   & 2.0           &  2.0   & 2.0  &  2.0   & 2.0\\
 
  Photon detection              &  1.0   & 1.5             &  1.0   & 1.5  &  1.0   & 1.5\\

   $\Lambda$  reconstruction         & 3.8  &  3.8           & 3.4  &  3.4  & 3.2  &  3.3 \\
  
  $\bar{\Lambda}$  reconstruction  & 0.9  &  0.9            & 0.4  &  0.4  & 0.2  &  0.2 \\

  $\etap$ mass window    &  0.6  &1.0 & 0.6  &1.0 & 0.6  &1.0\\
 
  Kinematic fit        & 5.2   & 5.3        &5.0    &4.2    &  4.8  & 4.7      \\

 Wrong photon        & 4.0  & -        &3.4    & -    &  1.4  & -      \\

 Background veto        & 1.1   & 0.1        &1.1    & 0.1    &  1.2  & 0.1      \\

Signal MC model        & 6.1  &5.4       &7.7    & 6.2    &  3.8  & 2.5      \\

  $\mathscr{B}(\etap\too X)$   & 1.4  &  1.3    & 1.4  &  1.3 & 1.4  & 1.3 \\

  $\mathscr{B}(\psi(3686)\too\gamma\chi_{cJ},~\Lambda\too p\pi^{-},~\bar{\Lambda}\too\bar{p}\pi^{+})$       &   \multicolumn{2}{c}{2.8}  &   \multicolumn{2}{c}{3.2} & \multicolumn{2}{c}{2.9}\\
  Fit  procedure   &   \multicolumn{2}{c}{6.3}  &   \multicolumn{2}{c}{8.5} & \multicolumn{2}{c}{5.7}\\
    Total number of $\psip$ events   &   \multicolumn{2}{c}{0.5}  &   \multicolumn{2}{c}{0.5} &   \multicolumn{2}{c}{0.5}    \\

  \hline
  Sum                  &   \multicolumn{2}{c}{11.9}            &   \multicolumn{2}{c}{10.6}   &  \multicolumn{2}{c}{9.7}\\
  \hline
  \hline
  \end{tabular}
\end{minipage}
\end{table*}

The uncertainty arising from the pion tracking is $1.0\%$ per track with the sample of $\psi(3686)\too\pi^{+}\pi^{-}J/\psi$, $J/\psi\too \ell^{+}\ell^{-}~(\ell=e,~\mu)$~\cite{track}, for pions from $\etap$ decays. The uncertainty of photon detection is $0.5\%$ per photon as determined by the $\EE\too\gamma\mu^{+}\mu^{-}$ process~\cite{photon}.

The combined efficiency, including tracking for proton/anti-proton and charged pion, and reconstruction for $\Lambda$ and $\bar{\Lambda}$, is studied by the control sample of $J/\psi\too p K^{-} \bar{\Lambda}+c.c.$~\cite{lambda_sys}. The relative efficiency discrepancy between data and MC simulation in each momentum and $|\!\cos\theta|$ bin is defined as  
 $\Delta\epsilon_{i}=\frac{\epsilon_{i}^{\rm data}}{\epsilon_{i}^{\rm MC}}-1$.
To quantify the systematic uncertainties, we apply a normalized weighting factor $w_{i}=\frac{\epsilon_{i}}{\epsilon_{\rm tot}}$, where $\epsilon_{i}$ is the MC efficiency in the $i-$th bin and $\epsilon_{\rm tot}$ is the total MC efficiency from our signal MC sample. The differences between data and MC simulation are subsequently evaluated by $\Sigma(w_{i}\times\Delta\epsilon_{i})$, which are assigned as the corresponding systematic uncertainties. 

To evaluate the systematic uncertainty associated with the $\etap$ mass window, we perform an unbined maximum likelihood fit to $M(\gamma\pi^{+}\pi^{-})$ (Mode I) and $M(\eta\pi^{+}\pi^{-})$ (Mode II) distributions in data. The signal shape of $\etap$ is modeled by an MC simulated shape convolved with a Gaussian function with fitted parameters. The smooth background is parameterized by a second-order Chebychev function. Using the best-fit signal shape parameters, we generate a toy MC sample to simulate the $\etap$ line shape. The systematic uncertainty is the difference in efficiency between the nominal and toy samples.

The uncertainty associated with the kinematic fit is estimated by correcting the helix parameters of charged tracks according to the method described in Ref.~\cite{kinematic}. The difference of the detection efficiencies obtained from MC samples with and without this correction is assigned as the uncertainty.

The systematic effect of possible photon mispairing is investigated with MC truth matching.  
Events in which the angle between the generated and reconstructed photon $\theta_{\text{match}}<7^{\circ}$ are considered correctly matched; otherwise they are counted as mismatched. The fractions of mismatched events are $4.0\%$, $3.4\%$, and $1.4\%$ for $\chi_{c0}$, $\chi_{c1}$, and $\chi_{c2}$, respectively.  
These fractions are assigned as systematic uncertainties.

The background veto critetia, include $M(\Lambda(\bar{\Lambda})\gamma)>1.2$ GeV/$c^{2}$, $|RM(\pi^{+}\pi^{-})-m_{J/\psi}|>10~{\rm MeV}/c^{2}$, $|RM(\gamma_{1}\gamma_{2})-m_{J/\psi}|>26~{\rm MeV}/c^{2}$, and $|M(\gamma_{1}\gamma_{2})-m_{\piz}|>14~{\rm MeV}/c^{2}$.
The uncertainty from the requirement of $M(\Lambda(\bar{\Lambda})\gamma)>1.2~{\rm GeV}/c^{2}$ is estimeted by the control sample of $\psi(3686)\too\pi^{+}\pi^{-}J/\psi,~J/\psi\too\Sigma^{0}\bar{\Sigma^{0}}$. The difference of the acceptance efficiencies between data and MC simulation is 0.01\% for $\Sigma^{0}$, and 0.1\% for $\bar{\Sigma^{0}}$, which are taken as the uncertainties. Similarly, the systematic uncertainty due to the criterion $|RM(\pi^{+}\pi^{-})-m_{J/\psi}|>10~{\rm MeV}/c^{2}$ is estimated by the control sample of $\psi(3686)\too\pi^{+}\pi^{-}J/\psi,~J/\psi\too\Lambda\bar{\Lambda}\eta$, the systematic uncertainty due to the criterion $|RM(\gamma_{1}\gamma_{2})-m_{J/\psi}|>26~{\rm MeV}/c^{2}$ is estimated by the control sample of $\psi(3686)\too\eta J/\psi,~\eta\too\gamma\gamma,~J/\psi\too\Lambda\bar{\Lambda}\pi^{+}\pi^{-}$, and the difference of the acceptance efficiencies between data and MC simulation is 0.2\% for each control sample, which is taken as the corresponding uncertainty.
As for the criterion $|M(\gamma_{1}\gamma_{2})-m_{\piz}|>14~{\rm MeV}/c^{2}$,
the photon energy resolution between data and MC simulation can affect the distribution of $M(\gamma_{1}\gamma_{2})$, we adjust the energy error in the reconstructed photon error matrix by 4\%~\cite{gamres}; the resulting change in acceptance is assigned as the uncertainty.

The uncertainties of the fit parameters from the 3-D fit will introduce a systematic uncertainty to the efficiency. This is estimated by generating 1000 parameter sets following a multivariate Gaussian distribution with the covariance matrix from the 3-D fit. Then we propagate these variations into the efficiency calculation. The variance of the resulting efficiency distribution is adopted as the systematic uncertainty from the signal MC model.

Systematic uncertainties from branching fractions are taken from the PDG~\cite{pdg}.

For the uncertainty of the signal shape, we modify the signal shape from the MC simulated shape to $({\rm BW}\cdot E_{\gamma}^3\cdot f_{d}(E_{\gamma}))\otimes {\rm Double ~Gaussian}$, where BW is a Breit-Wigner function, with its parameters are fixed at the values of $\chi_{cJ}$ from PDG~\cite{pdg}, $E_{\gamma}$ represents the energy of the M1 transition photon, $f_{d}(E_{\gamma})$ is the damping factor~\cite{dump}.
The double-Gaussian function models the detector resolution, whose parameters are derived from the corresponding MC simulation. 
The uncertainty from the background shape is evaluated by replacing the first-order Chebyshev polynomial with a second-order one.
The systematic uncertainty from the $\etap$ sideband regions is shifted by $\pm2$ MeV/$c^{2}$, and the maximum difference of final results is taken as the systematic uncertainty.
The normalization factors between the $\etap$ signal and sideband regions are determined from the fits to the $M(\gamma\pi^{+}\pi^{-})$ and $M(\eta\pi^{+}\pi^{-})$ distributions with all the parameters fixed to the best values. To consider the uncertainty from the $\etap$ sideband normalization factor, we free the value of the normalization factors, and the range of sideband scale factor is the nominal normalization factor $\pm1\sigma$, where $\sigma$ is the uncertainty of the sideband normalization factor from the fits to the $M(\gamma\pi^{+}\pi^{-})$ and $M(\eta\pi^{+}\pi^{-})$ distributions. For any individual systematic uncertainty related to the fit procedure, the difference in the final results is taken as the systematic uncertainty. When evaluating these systematic uncertainties, we also re-calculate the statistical significance. The smallest statistical significance obtained from these calculations is taken as the resulting significance that incorporates the systematic uncertainties, as presented in Table~\ref{tab:result}.

The total number of $\psi(3686)$ events in data is determined with the inclusive hadronic events to be $2.7124\times10^{9}$ with an uncertainty of 0.5\%~\cite{psipdata}.

The total systematic uncertainties are obtained by adding all systematic uncertainties in quadrature, assuming they are independent.

\section{SUMMARY}

Using a data sample of $(2.7124\pm0.014)\times10^{9}$ $\psi(3686)$ events collected with the BESIII detector at the BEPCII collider, we have studied the decays $\chi_{cJ}\rightarrow\Lambda\bar{\Lambda}\eta^{\prime}$ for $J=0$, $1$, $2$ via the radiative transition $\psi(3686)\rightarrow\gamma\chi_{cJ}$.  
The decays $\chi_{c0,2}\rightarrow\Lambda\bar{\Lambda}\eta^{\prime}$ are observed for the first time, and evidence for $\chi_{c1}\rightarrow\Lambda\bar{\Lambda}\eta^{\prime}$ is found.  
The measured branching fractions and significances are summarised in Table~\ref{tab:result}; they are significantly smaller than those for $\chi_{cJ}\rightarrow\Lambda\bar{\Lambda}\eta$~\cite{chicj1}, $\chi_{cJ}\rightarrow\Lambda\bar{\Lambda}\omega$~\cite{chicj2}, and $\chi_{cJ}\rightarrow\Sigma^{+}\bar{\Sigma}^{-}\eta$~\cite{chicj4}, which have larger phsp space.  
No excited baryon states or $\Lambda\bar{\Lambda}$ near-threshold enhancement are observed within the current statistical precision, consistent with previous measurements of $\chi_{cJ}\rightarrow B\bar{B}M$~\cite{chicj1, chicj2, chicj3, chicj4}.  
Further experimental studies of additional $\chi_{cJ}$ decay channels are essential to elucidate the properties of $\chi_{cJ}$ mesons, $B\bar{B}$ threshold enhancements, and possible excited baryon states.

\begin{acknowledgments}
The BESIII Collaboration thanks the staff of BEPCII (https://cstr.cn/31109.02.BEPC) and the IHEP computing center for their strong support. This work is supported in part by National Key R\&D Program of China under Contracts Nos. 2023YFA1606000, 2023YFA1606704; National Natural Science Foundation of China (NSFC) under Contracts Nos. 12375070, 11635010, 11935015, 11935016, 11935018, 12025502, 12035009, 12035013, 12061131003, 12192260, 12192261, 12192262, 12192263, 12192264, 12192265, 12221005, 12225509, 12235017, 12361141819; the Chinese Academy of Sciences (CAS) Large-Scale Scientific Facility Program; the Strategic Priority Research Program of Chinese Academy of Sciences under Contract No. XDA0480600; CAS under Contract No. YSBR-101; 100 Talents Program of CAS; Shanghai Leading Talent Program of Eastern Talent Plan under Contract No. JLH5913002; The Institute of Nuclear and Particle Physics (INPAC) and Shanghai Key Laboratory for Particle Physics and Cosmology; ERC under Contract No. 758462; German Research Foundation DFG under Contract No. FOR5327; Istituto Nazionale di Fisica Nucleare, Italy; Knut and Alice Wallenberg Foundation under Contracts Nos. 2021.0174, 2021.0299; Ministry of Development of Turkey under Contract No. DPT2006K-120470; National Research Foundation of Korea under Contract No. NRF-2022R1A2C1092335; National Science and Technology fund of Mongolia; Polish National Science Centre under Contract No. 2024/53/B/ST2/00975; STFC (United Kingdom); Swedish Research Council under Contract No. 2019.04595; U. S. Department of Energy under Contract No. DE-FG02-05ER41374.
\end{acknowledgments}


\begin{thebibliography}{**}

\bibitem{ee_chic1} M.~Ablikim \textit{et al.} (BESIII Collaboration), \href{https://journals.aps.org/prl/abstract/10.1103/PhysRevLett.129.122001}{\color{blue}Phys. Rev. Lett {\bf129}, 122001 (2022)}.


\bibitem{lam2omega} M.~Ablikim \textit{et al.} (BESIII Collaboration),  \href{https://journals.aps.org/prd/abstract/10.1103/PhysRevD.106.112011}{\color{blue}Phys. Rev. D {\bf106}, 112011 (2022).}

\bibitem{lam2phi} M.~Ablikim \textit{et al.} (BESIII Collaboration),  \href{https://journals.aps.org/prd/abstract/10.1103/PhysRevD.104.052006}{\color{blue}Phys. Rev. D {\bf104}, 152006 (2021).}

\bibitem{lam2eta} M.~Ablikim \textit{et al.} (BESIII Collaboration),  \href{https://journals.aps.org/prd/abstract/10.1103/PhysRevD.107.112011}{\color{blue}Phys. Rev. D {\bf107}, 112011 (2023).}

\bibitem{chicj1} M.~Ablikim \textit{et al.} (BESIII Collaboration),  \href{https://journals.aps.org/prd/abstract/10.1103/PhysRevD.106.072004}{\color{blue}Phys. Rev. D {\bf106}, 072004 (2022).}

\bibitem{chicj2} M.~Ablikim \textit{et al.} (BESIII Collaboration),  \href{https://journals.aps.org/prd/abstract/10.1103/PhysRevD.110.032022}{\color{blue}Phys. Rev. D {\bf110}, 032022 (2024).}

\bibitem{chicj3} M.~Ablikim \textit{et al.} (BESIII Collaboration), \href{https://journals.aps.org/prd/abstract/10.1103/PhysRevD.110.032016}{\color{blue}Phys. Rev. D {\bf110}, 032016 (2022).}

\bibitem{chicj4} M.~Ablikim \textit{et al.} (BESIII Collaboration), \href{https://journals.aps.org/prd/abstract/10.1103/PhysRevD.110.112013}{\color{blue}Phys. Rev. D {\bf110}, 112013 (2024).}

\bibitem{psipdata} M.~Ablikim \textit{et al.} (BESIII Collaboration), \href{https://iopscience.iop.org/article/10.1088/1674-1137/42/2/023001} {{\color{blue}Chin. Phys. C {\bf42}, 023001 (2013)}}; M.~Ablikim \textit{et al.} (BESIII Collaboration), \href{https://iopscience.iop.org/article/10.1088/1674-1137/ad595b} {{\color{blue}Chin. Phys. C {\bf48}, 093001 (2024)}}.

\bibitem{Ablikim:2009aa} M.~Ablikim \textit{et al.} (BESIII Collaboration), \href{https://doi.org/10.1016/j.nima.2009.12.050}{{\color{blue} Nucl. Instrum. Meth. A {\bf 614}, 345 (2010)}}.
    
\bibitem{Yu:IPAC2016-TUYA01} C.~H.~Yu \textit{et al.}, \emph{Proceedings of IPAC2016, Busan, Korea,} (2016), \href{https://doi.org/10.18429/JACoW-IPAC2016-TUYA01}{{\color{blue}10.18429/JACoW-IPAC2016-TUYA01.}}

 \bibitem{Ablikim:2019hff} M.~Ablikim \textit{et al.} (BESIII Collaboration),
   \href{https://iopscience.iop.org/article/10.1088/1674-1137/44/4/040001}{{\color{blue}Chin. Phys. C {\bf 44}, 040001 (2020)}}.

\bibitem{EcmsMea}
  J.~Lu, Y.~Xiao, X.~Ji,
  \href{https://doi.org/10.1007/s41605-020-00188-8}{{\color{blue}
  Radiat. Detect. Technol. Methods {\bf 4}, 337–344 (2020)}}.
 

\bibitem{EventFilter}
  J.~W.~Zhang, L.~H.~Wu, S.~S.~Sun \textit{et al.},
   \href{https://doi.org/10.1007/s41605-020-00188-8}{{\color{blue}
  Radiat. Detect. Technol. Methods {\bf 6}, 289–293 (2022)}}.


\bibitem{etof} X.~Li \textit{et al.}, \href{https://doi.org/10.1007/s41605-017-0014-2}{{\color{blue}Radiat. Detect. Technol. Methods {\bf 1}, 13 (2017)}}; Y.~X.~Guo \textit{et al.}, \href{https://doi.org/10.1007/s41605-017-0012-4}{{\color{blue}Radiat. Detect. Technol. Methods {\bf 1}, 15 (2017).}}
 
\bibitem{geant4} S.~Agostinelli \textit{et al.} (GEANT4 Collaboration),  \href{https://doi.org/10.1016/S0168-9002(03)01368-8}{{\color{blue} Nucl. Instrum. Meth. A {\bf 506}, 250 (2003).}}

\bibitem{KKMC} S.~Jadach, B.~F.~L.~Ward, and Z.~Was, \href{https://doi.org/10.1103/PhysRevD.63.113009} {{\color{blue}Phys. Rev. D {\bf 63}, 113009 (2001)}}; \href{https://doi.org/10.1016/S0010-4655(00)00048-5} {{\color{blue}Comput. Phys. Commun. {\bf 130}, 260 (2000).}}

\bibitem{ref:evtgen} D.~J.~Lange, \href{https://doi.org/10.1016/S0168-9002(01)00089-4}{ {\color{blue} Nucl. Instrum. Meth. A {\bf 462}, 152 (2001)}}; R.~G.~Ping, \href{https://doi.org/10.1088/1674-1137/32/8/001}{{\color{blue}Chin. Phys. C {\bf 32}, 599 (2008).}}


\bibitem{pdg} S.~Navas \textit{et al.} (Particle Data Group), \href{https://pdglive.lbl.gov/Viewer.action}{{\color{blue}Phys. Rev. D {\bf110}, 030001 (2024).}}

\bibitem{ref:lundcharm} J.~C.~Chen, G.~S.~Huang, X.~R.~Qi, D.~H.~Zhang, and Y.~S.~Zhu, \href{https://doi.org/10.1103/PhysRevD.62.034003}{{\color{blue} Phys  Rev. D {\bf 62}, 034003 (2000)}}; R.~L.~Yang, R.~G.~Ping, and H.~Chen, \href{https://doi.org/10.1088/0256-307X/31/6/061301}{ {\color{blue}Chin. Phys. Lett. {\bf 31}, 061301 (2014).}}

\bibitem{photos} E.~Richter-Was, \href{https://doi.org/10.1016/0370-2693(93)90062-M}{{\color{blue}Phys. Lett. B {\bf 303}, 163 (1993).}}

\bibitem{gammaangle} W.~Tanenbaum \textit{et al.}, \href{https://doi.org/10.1103/PhysRevD.17.1731} {{\color{blue}Phys. Rev. D {\bf 17}, 1731 (1978)}}; G.~R.~Liao, R.~G.~Ping, and Y.~X.~Yang, \href{https://doi.org/10.1088/0256-307X/26/5/051101} {{\color{blue}Chin. Phys. C {\bf26}, 051101 (2009)}}

\bibitem{etapgam2pi} M.~Ablikim \textit{et al.} (BESIII Collaboration), \href{https://journals.aps.org/prl/abstract/10.1103/PhysRevLett.120.242003}{\color{blue}Phys. Rev. Lett. \textbf{120}, 024003 (2018)}.

\bibitem{etapeta2pi} M.~Ablikim \textit{et al.} (BESIII Collaboration),  \href{https://journals.aps.org/prd/abstract/10.1103/PhysRevD.97.012003}{\color{blue}Phys. Rev. D. {\bf97}, 012003 (2018)}.

\bibitem{secondfit} M.~Xu \textit{et al.}, \href{https://iopscience.iop.org/article/10.1088/1674-1137/33/6/005}{\color{blue}
 Chin. Phys. C {\bf 33} 428 (2009)}.


\bibitem{track} M.~Ablikim \textit{et al.} (BESIII Collaboration), \href{https://journals.aps.org/prl/abstract/10.1103/PhysRevLett.105.261801}{\color{blue}Phys. Rev. Lett {\bf105}, 261801 (2010)}.

\bibitem{photon} M.~Ablikim \textit{et al.} (BESIII Collaboration), \href{https://doi.org/10.1007/JHEP08(2024)180}{\color{blue}JHEP {\bf 08}, 180 (2024)}.

\bibitem{lambda_sys} M.~Ablikim \textit{et al.} (BESIII Collaboration), \href{https://journals.aps.org/prd/abstract/10.1103/PhysRevD.108.112012}{\color{blue}Phys. Rev. D {\bf 108}, 112012 (2023)}.

\bibitem{kinematic} M.~Ablikim \textit{et al.} (BESIII Collaboration), \href{https://journals.aps.org/prd/abstract/10.1103/PhysRevD.87.012002}{\color{blue}Phys. Rev. D {\bf 87}, 012002 (2013)}.

\bibitem{gamres} M.~Ablikim \textit{et al.} (BESIII Collaboration), \href{https://journals.aps.org/prl/abstract/10.1103/PhysRevLett.104.132002}{\color{blue}Phys. Rev. Lett {\bf 104}, 132002 (2010).}

\bibitem{dump} V.~V.~Anashin \textit{et al.}, \href{https://www.worldscientific.com/doi/abs/10.1142/S2010194511000791}{\color{blue}Int. J. Mod. Phys. Conf. Ser. {\bf 02}, 188 (2011).}


\end{thebibliography}
\end{document}